\documentclass[journal]{IEEEtran}
\usepackage{graphicx,amssymb,amsmath}
\usepackage{multicol}
\usepackage[noadjust]{cite}
\usepackage{setspace}
\usepackage{stfloats}
\usepackage{midfloat}
\usepackage[normal]{threeparttable}
\usepackage{flushend,cuted}
\usepackage{cuted}
\usepackage{cases,subeqnarray}
\usepackage{bm,multirow,bigstrut}
\usepackage{textcomp}
\usepackage{latexsym,bm}
\usepackage{booktabs,changebar}
\usepackage{xcolor}
\usepackage{mathtools}

\IEEEoverridecommandlockouts
\begin{document}
%

%

\title{Effective Rate Analysis of MISO Systems over $\alpha$-$\mu$ Fading Channels
\thanks{This work was supported in part by the International Science \& Technology Cooperation Program of China (Grant No. 2015DFG12760),
the National Natural Science Foundation of China (Grant No. 61201185 and 61271266),  the Beijing Natural Science Foundation (Grant No. 4142027), the Foundation of Shenzhen government, and China Postdoctoral Science Foundation (No. 2014M560081).}}

\author{\authorblockN{Jiayi Zhang\authorrefmark{1}\authorrefmark{2}, Linglong Dai\authorrefmark{1}, Zhaocheng Wang\authorrefmark{1}, Derrick Wing Kwan Ng\authorrefmark{2}, and Wolfgang~H.~Gerstacker\authorrefmark{2}\\
\authorblockA{\authorrefmark{1}Tsinghua National Laboratory for Information Science and Technology (TNList)\\
Department of Electronic Engineering, Tsinghua University, Beijing 100084, P. R. China}\\
\authorblockA{\authorrefmark{2} Institute for Digital Communications, University of Erlangen-Nurnberg, D-91058 Erlangen, Germany}\\
 Email: \{jiayizhang, daill, zcwang\}@tsinghua.edu.cn, \{kwan, gersta\}@lnt.de}}

\maketitle

\begin{abstract}
The effective rate is an important performance metric of real-time applications in next generation wireless networks. In this paper, we present an analysis of the effective rate of multiple-input single-output (MISO) systems over $\alpha$-$\mu$ fading channels under a maximum delay constraint. More specifically, novel and highly accurate closed-form approximate expressions of the effective rate are derived for such systems assuming the generalized $\alpha$-$\mu$ channel model. In order to examine the impact of system and channel parameters on the effective rate, we also derive closed-form expressions of the effective rate in asymptotically high and low signal-to-noise ratio (SNR) regimes. Furthermore, connections between our derived results and existing results from the literature are revealed for the sake of completeness. Our results demonstrate that the effective rate is a monotonically increasing function of channel fading parameters $\alpha$ and $\mu$, as well as the number of transmit antennas, while it decreases to zero when the delay constraint becomes stringent.
\end{abstract}


\IEEEpeerreviewmaketitle
\section{Introduction}
Shannon's ergodic capacity of multiple-antenna systems has been extensively analyzed in pioneering works \cite{foschini1998limits,telatar1999capacity}. However, it cannot account for the quality of service (QoS) requirements of some emerging real-time applications for next-generation wireless networks, such as mobile video phone, interactive gaming, and multimedia streaming, where the tolerable delay is limited. Therefore, a novel performance metric is required to provision delay guarantees for such real-time applications. Motivated by this fact, Wu and Negi \cite{wu2003effective} proposed the concept of effective rate (or effective capacity, effective throughput) to consider the effect of statistical delay QoS guarantees for the achievable transmission rate. More specifically, the effective rate is defined as the maximum constant arrival rate at the transmitter when guaranteed statistical delay constraints can be satisfied.

Recently, the concept of effective rate has attracted many research interests for the analysis of single- and multiple-antenna communication systems \cite{femenias2009using,ahn2011throughput,cheng2013qos,Gursoy2011MIMO}. The effective rate of wireless networks was investigated for adaptive modulation and coding at the physical layer with an automatic repeat request protocol at the data-link layer in \cite{femenias2009using}. In \cite{ahn2011throughput}, an analytical model of the effective rate for proportional fair scheduling used in orthogonal frequency-division multiple access (OFDMA) systems was presented. For multiple-antenna systems, the authors in \cite{cheng2013qos} proposed the optimal power allocation scheme with statistical QoS provisioning to maximize the effective rate of virtual multiple-antenna wireless networks. The performance of multiple-antenna systems under QoS constraints is captured in \cite{Gursoy2011MIMO} through the effective rate formulation. Due to its low complexity and high performance, the MISO system is a promising technology for next-generation wireless networks. Therefore, a plethora of recent works focused on the effective rate of multiple-input single-output (MISO) systems, which cover the analysis of independent \cite{matthaiou2012analytic} and correlated \cite{zhong2011effective,zhong2012effective,guo2012performance} fading channels. However, these presented results are based on the assumption of a homogeneous scattering environment corresponding to Rayleigh, Rician, Nakagami-$m$, or generalized-$K$ fading models. In practice, the received signals reflected by a surface are spatially correlated, and the resulting signal is obtained in a nonhomogeneous scattering environment modeled by a nonlinear function of the modulus of the sum of the multipath components. To this end, the $\alpha$-$\mu$ fading channel is proposed in \cite{yacoub2007alpha} to better accommodate the statistical variations of the propagated signal in diverse field measurements of propagation environments \cite{chong2011wind, Michalopoulou2012body}. It has been proved that the $\alpha$-$\mu$ fading model includes Rayleigh, one-sided Gaussian, Weibull, exponential, Nakagami-$m$, and Gamma models as special cases.
Against this background, a basic understanding of the effects of nonhomogeneous $\alpha$-$\mu$ fading on the effective rate performance of MISO systems is provided in this paper. Only recently, we have derived novel expressions for two nonhomogeneous cases: $\eta$-$\mu$ and $\kappa$-$\mu$ fading channels \cite{zhang2013effective,Zhang2014effective}. To the best of authors' knowledge, however, there is no literature on the effective rate of MISO systems over $\alpha$-$\mu$ fading channels. The reason behinds this is twofold: First, the derivation of the exact expression for the probability density function (PDF) of the sum of $\alpha$-$\mu$ statistics is known to be very challenging; Second, significant mathematical challenges for the effective rate analysis are incurred when considering the complicated generalized $\alpha$-$\mu$ fading.

In this paper, we derive highly accurate closed-form expressions of the effective rate of MISO systems over independent and identical distributed (i.i.d.) $\alpha$-$\mu$ fading channels. Note that these expressions are given in the form of Meijer's-$G$ functions \cite{gradshtein2000table} and Fox's-$H$ functions \cite{carter1977distribution}, which can be easily evaluated and efficiently programmed in Mathematica. In order to intuitively investigate the impact of system and channel parameters on the effective rate, we also derive closed-form expressions of the effective rate in the asymptotically high and low signal-to-noise ratio (SNR) regime, respectively. Our results prove that the effective rate increases with larger value of fading parameters $\alpha$ and $\mu$, as well as more transmit antennas, while the tightened delay constraint will reduce the effective rate. For the sake of completeness, connections between our derived results and previously presented results are also provided in this paper.

The rest of the paper is organized as follows. Section \ref{se:model} describes the system model and the statistical characteristics of generalized $\alpha$-$\mu$ fading channels. In Section \ref{se:effective_rate}, we derive a novel expression for the exact effective rate of MISO systems over i.i.d. $\alpha$-$\mu$ fading channels. To quantify the effect of system and channel parameters on the effective rate, we also derive closed-form expressions both for high SNRs and for the minimum transmit energy per information bit. Finally, theoretical and simulation results are compared in Section \ref{se:numerical_results}, while Section \ref{se:conclusion} concludes the paper.

\emph{Notations:} Vectors and matrices are represented by lowercase bold typeface and uppercase bold typeface letters, respectively. The symbol ${\tt E}\{\cdot\}$ refers to the expectation operator of a random variable, $\Pr[\cdot]$ accounts for the probability, and $\textrm{tr}\left( \cdot \right)$ is the trace of a matrix. We use $(\cdot)^\dag$ to represent the Hermitian transpose, and the natural logarithm of a number is defined as $\ln(\cdot)$. Moreover, $\bf{I}$ denotes the unit matrix. Finally, $G[\cdot]$ represents the Meijer's-$G$ function \cite[Eq. (9.301)]{gradshtein2000table} and $H[\cdot]$ stands for the Fox's-$H$ function \cite[Eq. (2.1)]{carter1977distribution}.

\section{System and Channel Model}\label{se:model}
 \subsection{System Model}
We consider a point-to-point MISO system, where the transmitter consists of $N_t$ antennas and the receiver is deployed with single antenna. The input-output relationship over block fading channels can be expressed as
\begin{align}
y = {\bf{hx}} + n,
\end{align}
where ${\bf{h}} \in {\mathbb{C}^{1 \times {N_t}}}$ denotes the MISO channel's fading vector with elements ${\bf{h}} =\left[h_1,\cdots, h_k, \cdots, h_{N_t}\right]$, ${\bf{x}}\in {\mathbb{C}^{ {N_t}\times 1}}$ represents the transmit signal vector with the covariance ${\tt E}\{ {\bf{x}}{{\bf{x}}^\dag }\}  = {\bf{Q}}$ subject to the constraint $\textrm{tr}({\bf{Q}}) \le P$, where $P$ denotes the transmit power. Moreover, $n$ represents complex additive white Gaussian noise (AWGN) whose samples are i.i.d. zero-mean complex Gaussian random variables with variance $N_{0}$. Furthermore, we assume that the instantaneous channel state information is not available at the transmitter and equal power is allocated to each transmit antenna, i.e., ${\bf{Q}} = \frac{P}{{{N_t}}}{\bf{I}}$, so the transmit SNR is $\rho = \frac{P}{{{N_0}}}$.

\subsection{Effective Rate}
As introduced in \cite{wu2003effective}, we suppose that the data arrives in the buffer at a constant rate, and the service process of the transmitter is stationary. Then, the effective rate of the service process is defined as
\begin{equation}\label{eq:2}
\alpha \left( \theta  \right) =  -  \frac{1}{\theta T} \ln \left( {{\tt E}\left\{ {\exp \left( { - \theta TC} \right)} \right\}} \right),\; \; \; \theta  \ne 0,
\end{equation}
where $C$ represents the random throughput during a single block and $T$ denotes the block length. The delay QoS exponent $\theta$ is given by
\begin{equation}\label{eq:3}
\theta  =  - \mathop {\lim }\limits_{{l_\text{th}} \to \infty } \frac{{\ln \left(\Pr \left[ {L > {l_\text{th}}} \right]\right)}}{{l_\text{th}}},
\end{equation}
where ${l_\text{th}}$ denotes the specified threshold of queue length and $L$ is the equilibrium queue-length of the buffer assumed to be available at the transmitter. Note that we use the queue length as a QoS performance metric instead of delay to obtain an intuitive and tractable analysis. Moreover, the probability of ${L > {l_\text{th}}} $ decreases faster with a larger $\theta$. On the contrary, the system can tolerate an arbitrarily long delay when there is no delay constraint as $\theta  \rightarrow 0$. Capitalizing on such conditions, the effective rate coincides with the classic concept of Shannon's ergodic capacity.

In the case that the transmitter sends uncorrelated circularly symmetric zero-mean complex Gaussian signals, the effective rate of the MISO channel can be succinctly expressed as \cite{matthaiou2012analytic}
\begin{equation}\label{eq:4}
R\left( {\rho ,\theta } \right) =  - \frac{1}{A}{\log _2}\left( {{\tt E}\left\{ {{{\left( {1 + \frac{\rho }{{{N_t}}}{\bf{h}}{{\bf{h}}^\dag }} \right)}^{ - A}}} \right\}} \right) \;\;\textrm{bit/s/Hz},
\end{equation}
where $A \triangleq \frac{\theta TB}{\textrm{ln}2}$, with $B$ denoting the bandwidth of the system. In particular, $A$ represents a metric of delay constraint. It is clear from \eqref{eq:4} that the effective rate depends on the distribution of ${\bf{h}}{{\bf{h}}^\dag }$, which is analyzed in the following.
\newcounter{mytempeqncnt1}
\begin{figure*}[!b]
\normalsize
\setcounter{mytempeqncnt1}{\value{equation}}
\hrulefill
\vspace*{4pt}
\setcounter{equation}{9}
\begin{align}\label{eq:appro_meijer}
R\left( {\rho ,\theta } \right) =\frac{1}{A} - \frac{1}{A}{\log _2}\left( {\frac{{{\alpha} \sqrt k {l^{A - 1}}{{\left( {{N_t}\beta /\rho } \right)}^{{\alpha} {\mu} /2}}}}{{{{\left( {2\pi } \right)}^{l + k/2 - 3/2}}\Gamma \left( A \right)\Gamma \left( {\alpha}  \right)}}} \right) - \frac{1}{A}{\log _2}\left( {G_{l,k + l}^{k + l,l}\left[ {\frac{{{{\left( {{N_t}/\rho } \right)}^l}}}{{{{\left( {{\beta ^{{\alpha} /2}}k} \right)}^k}}}\left| {\begin{array}{*{20}{c}}
{\Delta \left( {l,1 - {\alpha} {\mu} /2} \right)}\\
{\Delta \left( {k,0} \right),\Delta \left( {l,A -{\alpha} {\mu} /2} \right)}
\end{array}} \right.} \right]} \right),
\end{align}
\setcounter{equation}{\value{mytempeqncnt1}}
\end{figure*}

\subsection{Sum of $\alpha$-$\mu$ RVs}
The $\alpha$-$\mu$ distribution is a general fading model, which accounts for the nonlinearity of a nonhomogeneous propagation environment. It consists of two parameters describing the physical properties of the fading, namely $\alpha$ and $\mu$. \textit{More specifically, the power parameter $\alpha$ represents the nonlinear function of the modulus of the sum of the multipath components, while the parameter $\mu$ is related to the number of multipath clusters.} For a single $\alpha$-$\mu$ fading link, the PDF of the instantaneous received SNR is given by \cite[Eq. (8)]{da2008highly}
\begin{equation}\label{eq:snr_single}
{f_{\gamma_1} }\left( {\gamma_1}  \right) = \frac{{{\alpha_1} {{\gamma_1} ^{{\alpha_1} {\mu_1} /2 - 1}}}}{{2{\beta_1 ^{{\alpha_1} {\mu_1} /2}}\Gamma \left( {\mu_1}  \right)}}\exp \left( { - {{\left( {\frac{{\gamma_1} }{\beta_1 }} \right)}^{{\alpha_1} /2}}} \right),
\end{equation}
where $\beta_1  \triangleq {\tt E}\left\{ \gamma_1  \right\}\Gamma \left({\mu_1}  \right)/\Gamma \left( {{\mu_1}  + 2/{\alpha_1} } \right)$, ${\tt E}\left\{ \gamma_1 \right\} = {{\hat r_1}^2} {{\Gamma \left( {{\mu_1}  + 2/{\alpha_1} } \right)}}/({{{{\mu_1} ^{2/{\alpha_1} }}\Gamma \left( {\mu_1}  \right)}})$, and $\hat r_1$ is defined as the ${\alpha_1}$-root mean value of the envelope random variable $R$, i.e., $\hat r_1 = \sqrt[{\alpha_1} ]{{E\left\{ {{R^{\alpha_1}}} \right\}}}$. Using \cite[Eq. (8.326.2)]{gradshtein2000table}, the $n$th moment of $\gamma_1$ can be expressed as
\begin{equation}\label{eq:snr_moment}
{\tt E}\left\{ {\gamma _1^n} \right\} = \frac{{{\beta_1 ^n}}}{{\Gamma \left( \mu_1  \right)}}\Gamma \left( {\mu_1  + \frac{{2n}}{\alpha_1 }} \right).
\end{equation}

Let $\gamma = \sum\nolimits_{i = 1}^{{N_t}} {{\gamma _i}} $ denote the sum of $N_t$ i.i.d. $\alpha$-$\mu$ branches instantaneous SNRs. It was shown in \cite{da2008highly} that a single $\alpha$-$\mu$ RV can exploited to approximate the sum of $\alpha$-$\mu$ RVs, and this approximation seems to achieve accurate results. More specifically, moment-based estimators are used to calculate the parameters $\alpha$, $\mu$, and $\hat r$ from the exact moments of $\gamma$. To this end, we need to derive the parameters $\alpha$, $\mu$, and $\hat r$ by solving the following nonlinear equations
\begin{align}\label{eq:moment_match}
\frac{{{{\tt E}^2}\left( \gamma \right)}}{{{{\tt E}}\left( {{\gamma^2}} \right) - {{\tt E}^2}\left( \gamma \right)}} &= \frac{{{\Gamma ^2}\left( {\mu  + 1/\alpha } \right)}}{{\Gamma \left( \mu  \right)\Gamma \left( {\mu  + 2/\alpha } \right) - {\Gamma ^2}\left( {\mu  + 1/\alpha } \right)}},\notag \\
\frac{{{{\tt E}^2}\left( {{\gamma^2}} \right)}}{{{{\tt E}}\left( {{\gamma^4}} \right) - {{{\tt E}}^2}\left( {{\gamma^2}} \right)}} &= \frac{{{\Gamma ^2}\left( {\mu  + 2/\alpha } \right)}}{{\Gamma \left( \mu  \right)\Gamma \left( {\mu  + 4/\alpha } \right) - {\Gamma ^2}\left( {\mu  + 2/\alpha } \right)}},\notag \\
\hat r &= \frac{{{\mu ^{1/\alpha }}\Gamma \left( \mu  \right){\tt E}\left( \gamma \right)}}{{\Gamma \left( {\mu  + 1/\alpha } \right)}},
\end{align}
where the moments in \eqref{eq:moment_match} can be evaluated by using the multinomial identity \cite[Eq. (10)]{peppas2011simple}:
\begin{align}\label{eq:multinomial_identity}
{\tt E}\left( {{\gamma^q}} \right) &= \sum\limits_{{j_1} = 0}^q {\cdots\sum\limits_{{j_{N_t - 1}} = 0}^{{j_{N_t - 2}}} {\left( {\begin{array}{*{20}{c}}
q\\
{{j_1}}
\end{array}} \right)} }\cdots\left( {\begin{array}{*{20}{c}}
{{j_{N_t - 2}}}\\
{{j_{N_t - 1}}}
\end{array}} \right)\notag \\
&  \times {\tt E}\left( {\gamma_{1}^{q - {j_1}}} \right)\cdots{\tt E}\left( {\gamma_{N_t}^{{j_{N_t - 1}}}} \right).
\end{align}
Note that an analytical solution to the system equations \eqref{eq:moment_match} is very difficult to obtain, while we can use numerical methods instead, such as the \emph{fsolve} function of Matlab and Maple. Having obtained these parameters, the approximated PDF expression of the sum can be written as
\begin{equation}\label{eq:snr_sum}
{f_{\gamma} }\left( {\gamma}  \right)\approx \frac{{{\alpha} {{\gamma} ^{{\alpha} {\mu} /2 - 1}}}}{{2{\beta ^{{\alpha} {\mu} /2}}\Gamma \left( {\mu}  \right)}}\exp \left( { - {{\left( {\frac{{\gamma} }{\beta }} \right)}^{{\alpha} /2}}} \right).
\end{equation}

\newcounter{mytempeqncnt}
\begin{figure*}[!b]
\normalsize
\setcounter{mytempeqncnt}{\value{equation}}
\hrulefill
\vspace*{4pt}
\setcounter{equation}{24}
\begin{align}
{\frac{{{E_b}}}{{{N_0}}}_{\min }}&= \frac{{\Gamma \left( \mu_1  \right)\ln 2}}{{\beta_1 \Gamma \left( {\mu_1  + 2/\alpha_1 } \right)}},\label{eq:Eb_N0_min}\\
{S_0}&=  \frac{{2{N_t}{\Gamma ^2}\left( {\mu  + 2/\alpha } \right)}}{{\left( {A + 1} \right)\left( {\Gamma \left( {\mu  + 4/\alpha } \right)\Gamma \left( \mu  \right) - {\Gamma ^2}\left( {\mu  + 2/\alpha } \right)} \right) + {N_t}{\Gamma ^2}\left( {\mu  + 2/\alpha } \right)}}.\label{eq:S0_min}
\end{align}
\setcounter{equation}{\value{mytempeqncnt}}
\end{figure*}

\section{Effective Rate over $\alpha$-$\mu$ Fading Channels}\label{se:effective_rate}
In this section, an exact analysis\footnote{The term \textit{exact analysis} is used to represent highly accurate approximations for the sake of distinguishing it from \textit{asymptotic analysis}.} of effective rate of MISO systems over i.i.d. $\alpha$-$\mu$ fading channels is presented. The asymptotically high and low-SNR regimes are also considered in order to provide physical insights into the impact of the system and channel parameters on the effective rate.

\subsection{Exact Analysis}
Substituting \eqref{eq:snr_sum} into \eqref{eq:4} and using \cite[Eq. (8.4.2.5)]{prudnikov1990integrals3}, \cite[Eq. (11)]{adamchik1990algorithm} and \cite[Eq. (21)]{adamchik1990algorithm}, we derive the exact effective rate of MISO systems over $\alpha$-$\mu$ fading channels as \eqref{eq:appro_meijer} at the bottom of this page, where $\Delta \left( {\epsilon, \tau } \right)=\frac{\tau }{\epsilon}, \frac{\tau+1 }{\epsilon},\cdots, \frac{\tau +\epsilon -1}{\epsilon}$, with $\tau$ being an arbitrary real value and $\epsilon$ a positive integer. Moreover, $l/k=\alpha/2$, where $l$ and $k$ are both positive integers. For example, if $\alpha= 0.8$, we should set $l=2$ and $k=5$. Furthermore, $l=\alpha$ and $k=2$ for the special case of $\alpha \in {\mathbb{Z}}^+$. Note that the evaluation of \eqref{eq:appro_meijer} can be done efficiently for special values of $\alpha$. However, for large values of $l$ and $k$, it is not efficient to compute \eqref{eq:appro_meijer}. Therefore, another method is adopted in the following to overcome this problem.
\setcounter{equation}{10}

We recall the well-known translation from the Meijer's $G$-function to the Fox's $H$-function as \cite[Eq. (8.3.2.21)]{prudnikov1990integrals3}
\begin{align}
H_{p,q}^{m,n}\left[ {x\left| {\begin{array}{*{20}{c}}
{\left[ {{a_p},1} \right]}\\
{\left[ {{b_p},1} \right]}
\end{array}} \right.} \right] = G_{p,q}^{m,n}\left[ {x\left| {\begin{array}{*{20}{c}}
{\left[ {{a_p}} \right]}\\
{\left[ {{b_p}} \right]}
\end{array}} \right.} \right].
\end{align}
With the help of \cite[Eq. (11)]{adamchik1990algorithm} and \cite[Eq. (21)]{adamchik1990algorithm}, the power and exponential functions can be expressed in terms of the Fox's $H$-function as
\begin{align}
{e^ { - x} } &= H_{0,1}^{1,0}\left[ {x\left| {\begin{array}{*{20}{c}}
 - \\
{\left( {0,1} \right)}
\end{array}} \right.} \right], \; \;\text{and}\label{eq:exp_H}\\
{\left( {1 + x} \right)^\omega } &= \frac{1}{{\Gamma \left( { - \omega } \right)}}H_{1,1}^{1,1}\left[ {x\left| {\begin{array}{*{20}{c}}
{\left( {\omega  + 1,1} \right)}\\
{\left( {0,1} \right)}
\end{array}} \right.} \right],\label{eq:power_H}
\end{align}
respectively. Substituting \eqref{eq:exp_H} and \eqref{eq:power_H} into \eqref{eq:4}, we get the Mellin--Barnes integral of the product of two Fox's $H$-functions. Then, the effective rate expression can be written as
\begin{align}\label{eq:appro_H}
&R\left( {\rho ,\theta } \right) = \frac{1}{A}\Bigg( 1 - {{\log }_2}\left( {\frac{\alpha }{{\Gamma \left( A \right)\Gamma \left( \mu  \right)}}} \right) \notag \\
&- {{\log }_2}\left( {H_{1,2}^{2,1}\left[ {{{\left( {\frac{{{N_t}}}{{\rho \beta }}} \right)}^{\alpha /2}}\left| {\begin{array}{*{20}{c}}
{\left( {1,\alpha /2} \right)}\\
{\left( {\mu ,1} \right),\left( {A,\alpha /2} \right)}
\end{array}} \right.} \right]} \right) \Bigg),
\end{align}
where we have used the integral identity \cite[Eq. (2.25.1.1)]{prudnikov1990integrals3} and the property \cite[Eq. (8.3.2.8)]{prudnikov1990integrals3} of Fox's $H$-functions. It is worth to mention that \eqref{eq:appro_H} is very compact which simplifies the mathematical algebraic manipulations encountered in the effective rate analysis.

For the case of Nakagami-$m$ fading channels, we set $\alpha=2$ and $\mu=m$. With the help of \cite[Eq. (07.33.26.0004.01)]{Wolfram2011function}, the effective rate expression of \eqref{eq:appro_meijer} reduces to
\begin{align}\label{eq:appro_meijer_naka}
R\left( {\rho ,\theta } \right) &= \frac{{m{N_t}}}{A}{\log _2}\left( {\frac{{\Omega \rho }}{{m{N_t}}}} \right) \notag \\
&- \frac{1}{A}{\log _2}\left( {U\left( {m{N_t};m{N_t} + 1 - A;\frac{{m{N_t}}}{{\Omega \rho }}} \right)} \right),
\end{align}
where $U(\cdot)$ is the Tricomi hypergeometric function \cite[Eq. (07.33.02.0001.01)]{Wolfram2011function} and $\Omega$ is the average gain of channel fading. Note that \eqref{eq:appro_meijer_naka} coincides with \cite[Eq. (6)]{matthaiou2012analytic}.

\subsection{Asymptotic Analysis}\label{se:effective_rate_asy}
In the high-SNR regime, we assume $\rho$ is large in the initial expression \eqref{eq:4} to obtain the following asymptotic effective rate
\begin{align}\label{eq:high_integral}
{R^\infty }\left( {\rho ,\theta } \right) &\approx  - \frac{1}{A}{\log _2}\Bigg( \frac{\alpha (\rho/N_t)^{-A} }{{2{\beta ^{\alpha \mu /2}}\Gamma \left( \mu  \right)}}\notag \\
&\times \int_0^\infty  {{\gamma ^{\alpha \mu /2 - A - 1}}\exp \left( { - {{\left( {\frac{\gamma }{\beta }} \right)}^{\alpha /2}}} \right)} d\gamma  \Bigg).
\end{align}
By invoking \cite[Eq. (3.326.2)]{gradshtein2000table} and after some straightforward algebraic manipulations, we can derive a tractable result as
\begin{align}\label{eq:high}
{R^\infty }\left( {\rho ,\theta } \right) \approx  {\log _2}\left( {\frac{{\beta \rho }}{{{N_t}}}} \right) - \frac{1}{A}{\log _2}\left( {\frac{{\Gamma \left( {\mu  - 2A/\alpha } \right)}}{{\Gamma \left( \mu  \right)}}} \right).
\end{align}
Note that the condition required for using \cite[Eq. (3.326.2)]{gradshtein2000table} for the integral in \eqref{eq:high_integral} is satisfied in our calculation by taking $A<\alpha \mu /2-1$. The above result indicates that the high-SNR slope is $S_\infty=1$, which is independent of $\beta$. The same observations were made for the Rayleigh, Rician, and Nakagami-$m$ cases \cite{matthaiou2012analytic,zhong2011effective}.

We now investigate the effective rate in the low-SNR regime. Intuitively, a second-order Taylor expansion for $\rho \rightarrow 0$ can approximate the low-SNR effective rate. However, the authors in \cite{lozano2003multiple} proved that the Taylor expansion method may in fact result in misleading conclusions regarding the impact of the channel in the low SNR regime. Hence, it is beneficial to explore the effective throughput at low SNRs as \cite{lozano2003multiple}
\begin{align}\label{eq:19}
R\left( {\frac{{{E_b}}}{{{N_0}}}}, \theta \right) \approx {S_0}{\log _2}\left( {{{\frac{{{E_b}}}{{{N_0}}}}}/{{{{\frac{{{E_b}}}{{{N_0}}}}_{\min }}}}} \right).
\end{align}
The normalized transmit energy per information bit $\frac{E_b}{N_0}_\text{min}$ and the wideband slope $S_0$ can be respectively formulated as
\begin{align}
{\frac{{{E_b}}}{{{N_0}}}_{\min }} &\buildrel \Delta \over = \mathop {\lim }\limits_{\rho  \to 0} \frac{\rho }{{R\left( {\rho ,\theta } \right)}} = \frac{1}{{R'\left( {0,\theta } \right)}},\label{eq:Eb_N0}\\
{S_0} &\buildrel \Delta \over =  - \frac{{2{{[R'\left( {0,\theta } \right)]}^2}\ln 2}}{{R''\left( {0,\theta } \right)}},\label{eq:S0}
\end{align}
where $R'\left( {0,\theta } \right)$ and $R''\left( {0,\theta } \right)$ represent the first and second order derivatives of $R\left( {0,\theta } \right)$ with respect to $\theta$, which can be expressed as
\begin{align}
R'\left( {0,\theta } \right) &= \frac{1}{{{N_t}\ln 2}}{\tt E}\left\{ {{\bf{h}}{{\bf{h}}^\dag }} \right\},\label{eq:24}\\
R''\left( {0,\theta } \right) &= \frac{A}{{N_t^2\ln 2}}{\Big( {{\tt E}\left\{ {{\bf{h}}{{\bf{h}}^\dag }} \right\}} \Big)^2} - \frac{{A + 1}}{{N_t^2\ln 2}}{\tt E}\left\{ {{{\left( {{\bf{h}}{{\bf{h}}^\dag }} \right)}^2}} \right\}. \label{eq:25}
\end{align}
Recall that for i.i.d. MISO $\alpha$-$\mu$ fading channels,
\begin{align}\label{eq:26}
{\tt E}\left\{ {{\bf{h}}{{\bf{h}}^\dag }} \right\} = \sum\limits_{k = 1}^{{N_t}} {{\tt E}\left\{ {{{\left| {{h_k}} \right|}^2}} \right\}}    = {N_t}\beta_1 \frac{{\Gamma \left( {\mu_1  + 2/\alpha_1 } \right)}}{{\Gamma \left( \mu_1  \right)}}.
\end{align}
Using an approach similar to Appendix I of \cite{zhong2011effective}, we can derive
\begin{align}\label{eq:45}
&{\tt E}\left\{ {{{\left( {{\bf{h}}{{\bf{h}}^\dag }} \right)}^2}} \right\}= \sum\limits_{k = 1}^{{N_t}} {{\tt E}\left\{ {{{\left| {{h_k}} \right|}^4}} \right\}}  + \sum\limits_{k = 1}^{{N_t}} {\sum\limits_{j = 1,j \ne k}^{{N_t}} {{\tt E}\left\{ {{{\left| {{h_k}} \right|}^2}{{\left| {{h_j}} \right|}^2}} \right\}} }  \notag \\
& = \frac{{{N_t}{\beta ^2}}}{{\Gamma \left( \mu  \right)}}\left( {\Gamma \left( {\mu  + \frac{4}{\alpha }} \right) + \left( {{N_t} - 1} \right)\frac{{{\Gamma ^2}\left( {\mu  + 2/\alpha } \right)}}{{\Gamma \left( \mu  \right)}}} \right).
\end{align}
Substituting \eqref{eq:26}, \eqref{eq:45}, \eqref{eq:24}, and \eqref{eq:25} into \eqref{eq:Eb_N0} and \eqref{eq:S0}, the minimum $\frac{E_b}{N_0}$ and the wideband slope $S_0$ are respectively given by \eqref{eq:Eb_N0_min} and \eqref{eq:S0_min} at the bottom of this page.
\setcounter{equation}{25}

It is interesting to observe that the minimum $\frac{E_b}{N_0}$ is independent of the delay constraint $A$, whereas the wideband slope $S_0$ is independent of $\beta$. Moreover, $S_0$ is a decreasing function in $A$, while it is a monotonically increasing function in $N_t$. These results prove that a tighter delay constraint $A$ decreases the effective rate and an increasing number of antennas will yield a higher effective rate, respectively. The maximum value of the wideband slope is $S_0=2$ when taking $\mu \rightarrow \infty$ or $N_t \rightarrow \infty$. Note that for the case of Nakagami-$m$ fading channels ($\mu=m$ and $\alpha=2$), the minimum $\frac{E_b}{N_0}$ in \eqref{eq:Eb_N0_min} and wideband slope $S_0$ in \eqref{eq:S0_min} reduce to
\setcounter{equation}{26}
\begin{align}
{\frac{{{E_b}}}{{{N_0}}}_{\min }}   &= \frac{ \ln 2}{\Omega}, \;\;\text{and}\label{eq:Eb_N0_min_naka}\\
{S_0}   &=  \frac{2 m {N_t}} {A+1+m {N_t}},\label{eq:S0_min_naka}
\end{align}
respectively, which is in agreement with \cite[Eq. (20)-(21)]{gradshtein2000table}.

\section{Numerical Results}\label{se:numerical_results}
In this section, we verify the analytical results presented in Section \ref{se:effective_rate} by computer simulations, and use them to study the effective rate performance of MISO systems over $\alpha$-$\mu$ fading channels.  Without loss of generality, the bandwidth of the system is normalized to $B=1$ Hz. The simulation results are derived by averaging the results over $10^7$ i.i.d. $\alpha$-$\mu$ channel realizations with unit power, which are generated by the sum of independent Gaussian RVs method given in \cite{yacoub2007alpha}. These results provide meaningful insights regarding the impact of different system and channel parameters on the effective rate of MISO systems transmitting over $\alpha$-$\mu$ fading channels.

\begin{figure}[!t]
\centering
\includegraphics[scale=0.5]{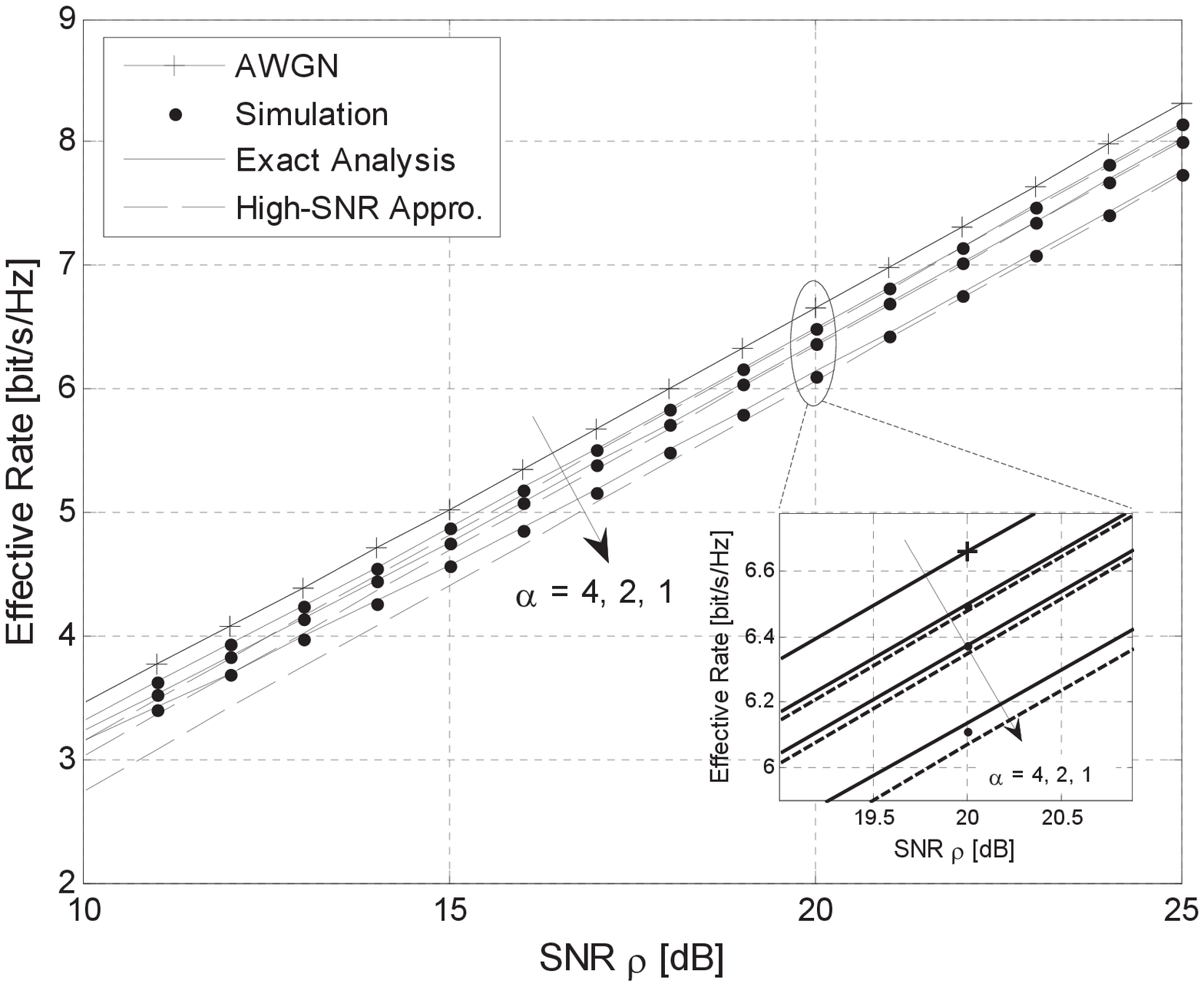}\\
\caption{Simulated and high-SNR approximate effective rate against the transmit SNR $\rho$ for MISO systems over i.i.d. $\alpha$-$\mu$ fading channels ($N_t = 2$, $A = 0.5$, and $\mu = 2$). Moreover, the effective rate of the non-fading channel, i.e., AWGN channel, is plotted to serve as a benchmark for comparison.
\label{High_alpha}}
\end{figure}

\begin{figure}[!t]
\centering
\includegraphics[scale=0.5]{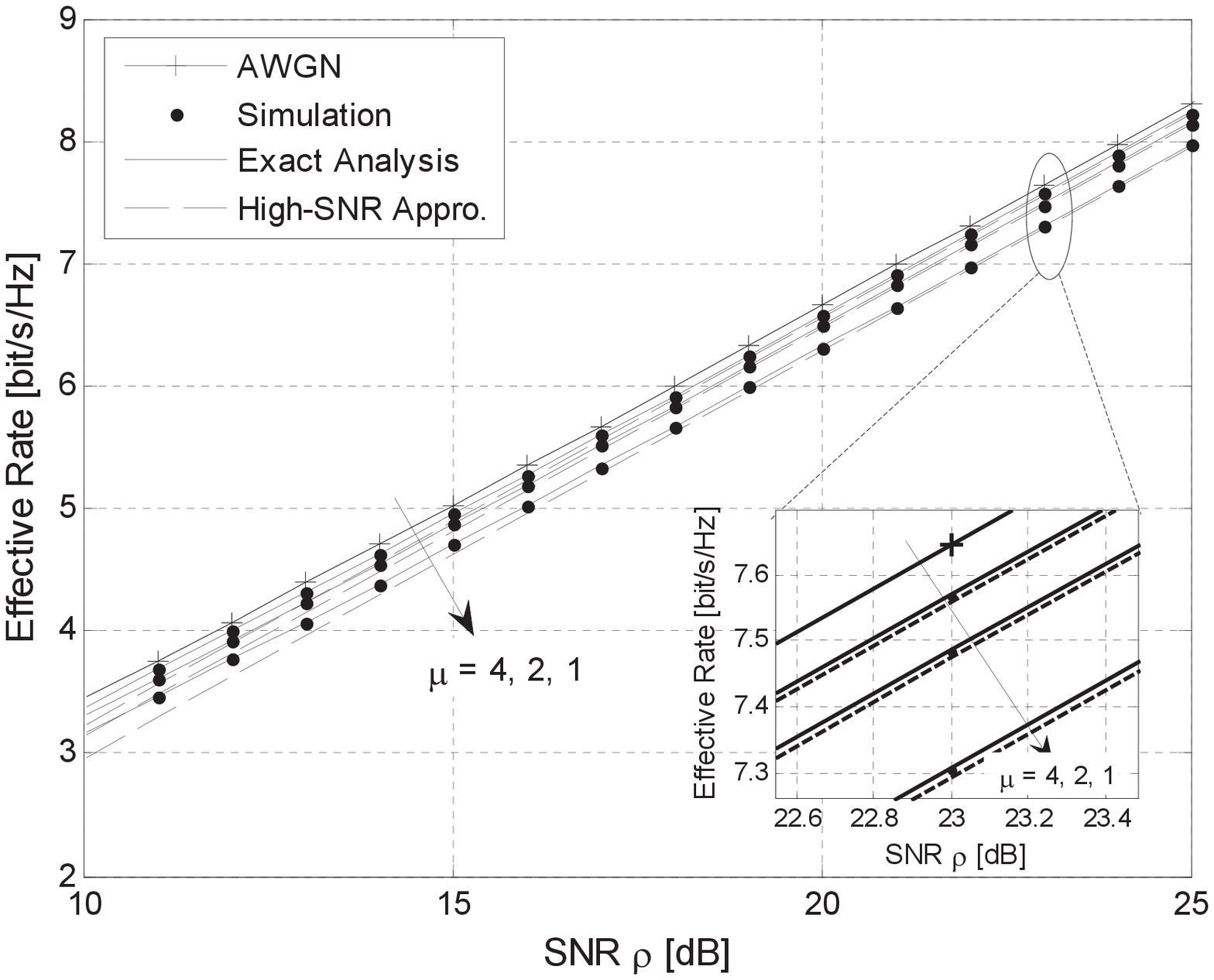}\\
\caption{Simulated and high-SNR approximate effective rate against the transmit SNR $\rho$ for MISO systems over i.i.d. $\alpha$-$\mu$ fading channels ($N_t = 2$, $A = 0.5$, and $\alpha = 4$). Moreover, the effective rate of the non-fading channel, i.e., AWGN channel, is plotted to serve as a benchmark for comparison.
\label{High_mu}}
\end{figure}

In Fig. \ref{High_alpha} and Fig. \ref{High_mu}, we consider the effects of fading parameters ($\alpha$ and $\mu$) on the effective rate of MISO systems. More specifically, the effective rate results of our computer simulations are compared to that of the exact and high-SNR approximate analytical expressions provided in (\ref{eq:appro_meijer}), (\ref{eq:appro_H}), and \eqref{eq:high} for different transmit SNRs. For comparison purposes, we also show the effective rate of MISO systems over AWGN channels as a benchmark. It is easy to see that the exact analytical expression of the effective rate is very accurate in the entire SNR regime. Besides, the high-SNR approximation is quite tight even in moderate SNRs and its accuracy is improved for larger values of the fading parameters, which implies that it can efficiently predict the effective rate over a wide range of SNR. Moreover, the high-SNR results show that the high-SNR slope $S_\infty$ is 1, which confirms our analysis in Section \ref{se:effective_rate_asy}. Note that a performance improvement of the effective rate is observed as $\alpha$ and $\mu$ increases. This observation is anticipated, since a large value of $\mu$ results in more multipath components and a large value of $\alpha$ accounts for a larger fading gain from the physical model, respectively. We also note that the gap between the AWGN and fading channel curves becomes smaller as $\alpha$ and $\mu$ get larger. Finally, it is worth mentioning that the effects of fading parameters on the effective rate become less pronounced as they increase.

\begin{figure}[!t]
\centering
\includegraphics[scale=0.5]{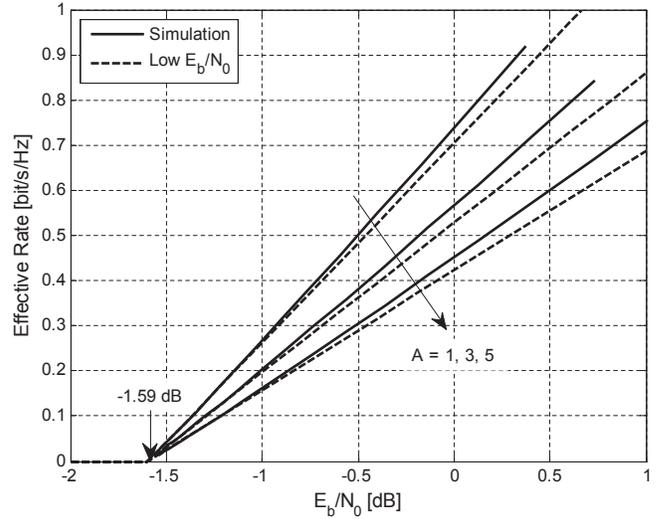}\\
\caption{Simulated, low-$E_b/N_0$ approximate effective rate against the $E_b/N_0$ for MISO systems over i.i.d. $\alpha$-$\mu$ fading channels ($N_t = 2$,  $\alpha = 2$, and $\mu = 2$).
\label{low}}
\end{figure}

The simulated and low-$E_b/N_0$ approximate effective rate \eqref{eq:19} are depicted against the transmit energy per bit $E_b/N_0$ for different delay constraints $A$ in Fig. \ref{low}. Clearly, the low-$E_b/N_0$ approximations are sufficiently tight and become exact at low $E_b/N_0$ values for all the considered scenarios. The effective rate is a monotonically decreasing function of $A$, which implies that tightening the delay constraints reduces the effective rate. However, the change of the delay constraint $A$ does not affect the minimum $E_b/N_0$, which is $-1.59$ dB in our case. In addition, the curves in Fig. \ref{low} also show that the accuracy of low-$E_b/N_0$ approximate solution is improved for smaller values of the delay constraints.

\section{Conclusions}\label{se:conclusion}
In this paper, we have presented a novel approach to analyze the effective rate of MISO systems over i.i.d. $\alpha$-$\mu$ fading channels. The proposed technique is based on the highly accurate approximation of the sum of i.i.d. $\alpha$-$\mu$ RVs by another $\alpha$-$\mu$ RV. Novel and analytical expressions of the exact effective rate of MISO systems over i.i.d. $\alpha$-$\mu$ fading channels have been derived. Moreover, we have presented closed-form expressions of the effective rate in the high-SNR regime to gain physical insights into the impact of system and channel parameters on the effective rate performance. For example, the effective rate can be improved by utilizing more transmit antennas as well as in a propagation environment with larger values of $\alpha$ and $\mu$. In addition, our analysis provides the minimum required transmit energy per information bit for reliably conveying any non-zero rate at low SNRs. Finally, numerical results corroborate the high accuracy of the proposed approximations. Our analytical results serve as a performance benchmark for our future work on the performance analysis of the multi-user scenario.


\bibliographystyle{IEEEtran}
\bibliography{IEEEabrv,Ref}

\end{document}